\documentclass[superscriptaddress,prl,twocolumn,nofootinbib,showpacs,preprintnumbers]{revtex4}
\usepackage{graphicx,epsfig}
\begin{document}

\title{Orbifold Unification for the Gauge and Higgs Fields and Their Couplings}

\author{Ilia Gogoladze}
\affiliation{Department of Physics, University of Notre Dame, 
Notre Dame, IN 46556, USA}

\author{Tianjun Li}
\affiliation{School of Natural Sciences, Institute for Advanced Study,
  Einstein Drive, Princeton, NJ 08540, USA}

\author{Yukihiro Mimura}
\affiliation{Department of Physics, University of Regina, 
Regina, Saskatchewan S4S 0A2, Canada}

\author{S. Nandi} 
\affiliation{Department of Physics, Oklahoma State University, 
Stillwater, OK 74078-3072, USA}



\begin{abstract}

We present an orbifold GUT model in which the Higgs trilinear 
couplings are unified with the three Standard Model gauge couplings. 
The model is constructed as an $N=2$ supersymmetric $SU(8)$ gauge theory 
in six dimensions, which is reduced to a supersymmetric Standard Model with
three singlets and extra $U(1)$ 
factors upon compactification. Such an unification is in good agreement 
with experiments. The predicted upper limit for the lightest CP-even 
neutral Higgs boson is somewhat larger than in the MSSM, and can be 
tested in the upcoming Large Hadron Collider.

\end{abstract}

\pacs{11.25.Mj, 12.10.Kt, 12.60.Fr}

\preprint{OSU-HEP-05-02}

\maketitle

{\em Introduction --}  
The Minimal Supersymmetric Standard Model (MSSM) is the most
natural extension of 
the Standard Model (SM).
It elegantly solves the gauge hierarchy problem, contains
neutralino as the cold dark matter candidate, and naturally
accommodates the gauge coupling unification \cite{Langacker:1990jh,Amaldi:1991cn}. 
Depending on the supersymmetry breaking mechanism, it also has
distinct predictions for the sparticles' (supersymmetric partners
of the SM particles) spectra which can be tested at the 
upcoming colliders such as the Large Hadron Collider (LHC)
and the future International Linear Collider (ILC). Despite
all these successes, there are several unanswered questions
within the MSSM. Why the bilinear supersymmetric Higgs
mass $\mu$ (in the superpotential) involving the up and down
Higgs superfields, $\mu H_u H_d$,
is at the TeV scale but not at the Planck scale? 
This is known as the $\mu$ problem. 
Also the predicted upper bound for the mass of the 
lightest CP-even neutral Higgs boson $h^0$ is around
130 GeV \cite{Okada:1990vk}, which is not much higher than
 the current experimental lower limit, 114 GeV. 
Moreover, the prediction for the proton decay
rate through dimension-5 operators is uncomfortably close to
the current experimental bounds. 

 These problems in the MSSM have prompted many to consider
the possible extensions 
to the Next to the Minimal Supersymmetric
Standard Model (NMSSM) \cite{Nilles:1982dy} by the addition of one or more Higgs
singlets to the usual two doublets present in the MSSM. Such
extensions can be used to solve the $\mu$ problem, and extend
the upper mass limit for the Higgs mass. Additionally, in an orbifold
GUT model (such as the one being proposed here), the doublet-triplet
splitting problem is naturally solved, thus avoiding the possible
problem with the proton decay rate~\cite{Orbifold}.
With the addition of one singlet
Higgs field, we can have the additional trilinear interaction terms
$\lambda S H_u H_d$ and $\kappa S^3/3$ in the superpotential. After
minimization of the scalar Higgs potential,
 $S$ can obtain a vacuum expectation value (VEV)
around the supersymmetry-breaking scale ($M_{SUSY}$), and generate
an effective $\mu$ term with $\mu = \lambda \langle S \rangle$.
 Thus, the $\mu$ problem is solved. However,
such NMSSM lacks definite predictions because 
the Higgs couplings $\lambda$ and $\kappa$ are 
completely arbitrary. Is there a theoretical framework
in which the values of $\lambda$ and $\kappa$
get determined?

In this work, we present a supersymmetric Standard Model with three
SM  singlets and following superpotential
\begin{equation}
W=\lambda H_uH_dS-\kappa SS_1S_2 \,,
\label{mm-SSM}
\end{equation}
where the Yukawa couplings 
$\lambda$ and $\kappa$ get determined in terms of the gauge
couplings, and thus making this model predictive and
testable by experiment. The idea is simple, and very attractive,
and has lead to the unification of gauge and Yukawa couplings \cite{GMS}.
%
%
We use the framework of extra dimensions with supersymmetry.
The two Higgs doublets, as well as the singlets are all part
of the gauge multiplet in higher dimensions, and the non-minimal
interactions involving the $\lambda$ and $\kappa$ are just part of
the gauge interactions in higher dimensions. We present the realization
of this idea below.

{\em Formalism and the Model --}
We consider a theory with a gauge symmetry $G$ in six dimensions (6D)
with $N=2$ supersymmetry. (The two extra dimensions will be
compactified on a suitable orbifold such that the gauge symmetry
is broken down to the SM with possibly some extra $U(1)$ factors,
and the supersymmetry is broken down to $N=1$). The $N=2$ supersymmetry
in 6D corresponds to $N=4$ supersymmetry in 4D, and thus only the
gauge multiplet can be introduced in the bulk. In terms of 4D $N=1$
language, the six dimensional gauge multiplet contains a vector
multiplet, $V$, and three chiral multiplets $\Sigma_1$, $\Sigma_2$, and 
$\Sigma_3$ in the adjoint representation of the gauge group $G$. 
The bulk action \cite{NMASWS},
written in 4D $N=1$ language and in the Wess-Zumino gauge, contains
the following trilinear term of the chiral multiplets
\begin{equation}
{\cal S} = \int d^6 x \int d^2 \theta \frac1{kg^2} {\rm Tr} 
\left( - {\sqrt 2} \Sigma_1 [\Sigma_2,\Sigma_3]\right)
+ {\rm H.C.} \,,
\label{TriTerm}
\end{equation}
where $k$ is the normalization factor for the group generators.
If the SM singlet Higgs field, and the up- and down-type Higgs doublets are contained
in the zero modes of the chiral multiplets $\Sigma_1$, $\Sigma_2$, and 
$\Sigma_3$, the gauge interaction term, Eq.~(\ref{TriTerm}), includes the trilinear
Higgs interaction term $\lambda S H_u H_d$ with the coupling $\lambda$
determined in terms of the gauge coupling $g$. In this construction, the 
singlet Higgs field, the two Higgs doublets, and the gauge fields are all unified
in a single multiplet of the gauge symmetry group $G$ in higher dimensions.
In the NMSSM, we also need a cubic term, $\kappa S^3/3$ for the singlet field
$S$ to develop a VEV. We can see from Eq.~(1) that we need three SM singlet Higgs
fields to be present in the zero modes of $\Sigma_1$, $\Sigma_2$, and 
$\Sigma_3$ leading to a trilinear term $\kappa S S_1 S_2$.

We now address what bulk gauge symmetry we need to unify both $\lambda$
and $\kappa$ with the gauge couplings. To obtain both the Higgs doublets
and the singlets as zero modes in 4D from the extra dimensional components 
of the higher dimensional gauge multiplet, and also to break the supersymmetry to $N=1$,
the minimal bulk gauge symmetry needed is 
$SU(4)_W$. In this case, $SU(4)_W$ is broken down to 
$SU(2)_L \times U(1)_Y \times U(1)^\prime$ upon compactification, and the
adjoint {\bf 15}-dimensional representation will have two doublets, $H_u$
and $H_d$, and a singlet $S$ as the zero modes. In this case, we can obtain
only the trilinear interaction $\lambda S H_u H_d$ from the bulk gauge
interaction with $\lambda=g_2$, where $g_2$ is the weak gauge coupling.
The minimal gauge symmetry in the bulk to include both the $\lambda$ and 
$\kappa$ terms from the zero mode bulk interaction is $SU(5)_W$. The $SU(5)_W$
gauge symmetry in the bulk, upon compactification to 4D, is broken down to 
$SU(2)_L \times U(1)_Y \times U(1)^\prime \times U(1)''$. The $SU(5)_W$ adjoint
representation, $\bf 24$, decomposed under the 
$SU(2)_L \times U(1)_Y \times U(1)^\prime \times U(1)''$
contains the two Higgs doublets, $H_u$ and $H_d$, as well as three singlets 
 $S$, $S_1$ and $S_2$ as zero modes. The bulk gauge interaction contains the 
 $\lambda S H_u H_d$, as well as $\kappa S S_1 S_2$ terms, giving rise to 
 $\lambda =\kappa = g_2$ at the compactification scale. With $SU(3)_C \times SU(5)_W$
 as the gauge symmetry in the bulk, we can include color interaction, but
 this does not unify the three SM gauge couplings. Thus we are naturally lead to an 
 $SU(8)$ gauge symmetry in the bulk to unify all three SM gauge couplings
 with $\lambda$ and $\kappa$.
 
 The model we propose for the gauge and Higgs trilinear coupling
unification is in six dimensions
 with $N=2$ supersymmetry, and $SU(8)$ gauge symmetry. The two extra dimensions 
 $x_5$ and $x_6$ are compactified on a $T^2/Z_6$ orbifold,
 which is obtained from torus $T^2$ by moduloing the $Z_6$ 
equivalent class: $z \sim \omega z$,
where $z$ is the complex coordinate of the extra dimensions and $\omega = e^{i \pi/3}$.
%
The transformation property for the vector multiplet, $V$ is
\begin{eqnarray}
 V(x^{\mu}, ~\omega z, ~\omega^{-1} {\bar z}) &=& R \,
 V(x^{\mu}, ~z, ~{\bar z}) R^{-1}~,~\, 
\label{S3trans}
\end{eqnarray}
where $R$ is an $8\times8$ matrix and $R^6=I$.
The transformation rules for the three chiral multiplets $\Sigma_1$, $\Sigma_2$, and 
$\Sigma_3$ are obtained by multiplying the right hand side of the Eq.~(\ref{S3trans}) by
the additional factors $\omega^{-1}$, $\omega^{-1-m}$, and $\omega^{2+m}$
respectively, where $m$ is an integer. 
These transformations keep the bulk action invariant
and non-trivial $R$ breaks the bulk gauge symmetry $G$ at the 4D fixed point~\cite{Li:2003ee}.
We choose the matrix $R$ to be
\begin{eqnarray}
R &=& {\rm diag} \left(+1, +1, +1,
 \omega^{n_1}, \omega^{n_1}, \omega^{n_2}, \omega^{n_3}, \omega^{n_4} \right)\,.
\end{eqnarray}
Then, for unequal values of the integers $n_1$, $n_2$, $n_3$ and $n_4$, upon 
compactification to 4D, the $SU(8)$
gauge symmetry breaks to $SU(3)_C\times SU(2)_L\times U(1)_Y \times U(1)_{\alpha}
\times U(1)_{\beta} \times U(1)_{\gamma}$, and the $N=4$ supersymmetry in 4D is broken
down to $N=1$ by an appropriate choice of $m$.
For the choice of $m=1$, and $n_1=5, n_2=4, n_3=2$ and $n_4=1$, the
zero modes of the {\bf 63}-dimensional vector multiplet are the gauge bosons (and gauginos) corresponding 
to the unbroken gauge symmetry, while the zero modes of the three {\bf 63}-dimensional
chiral multiplets $\Sigma_1$, $\Sigma_2$ and $\Sigma_3$, and their
quantum numbers are given in Table \ref{Spectrum0}. These zero modes include the Higgs bosons of 
above model in the compactified 4D theory. From the bulk action in Eq.~(\ref{TriTerm}), we
obtain the non-minimal Higgs interactions for the zero modes
of the kinetic-normalized chiral multiplets 
%
\begin{eqnarray}
{\cal S} &=& \int d^6 x \left[ \int d^2 \theta \ g_6 \left(
S H_u H_d - S S_1 S_2 - Q_X \overline{D}_X H_u 
\right. \right. \nonumber\\ && \left. 
+S_2 H_u' H_d' 
- S_1' H_u' H_d 
+ S_1' D_{\delta}
\overline{D}_{\delta} \right)
 + {\rm H. C.}\bigg]\,,
\end{eqnarray}
where $g_6$ is the 6D gauge coupling whose mass dimension is $-1$.

\begin{table}[t]
\caption{The zero modes of the chiral multiplets 
$\Sigma_1$, $\Sigma_2$ and $\Sigma_3$, and
their quantum numbers under the $SU(3)_C\times SU(2)_L\times U(1)_Y \times U(1)_{\alpha}
\times U(1)_{\beta} \times U(1)_{\gamma}$ gauge symmetry.
The anti-symmetric subscripts $Qij$
($Qij=-Qji$) are the charges under
the $U(1)_Y \times U(1)_{\alpha}
\times U(1)_{\beta} \times U(1)_{\gamma}$ gauge symmetry:
$Q12=({-{5/6}}, {0}, {0}, {0})$,
$Q13=({-{1/3}}, {8/15}, {-{1/2}}, {-{1/2}})$,
$Q14=({-{1/3}}, {{8/15}}, {{1/2}}, {-{1/2}})$,
$Q15=({-{1/3}}, {{8/{15}}}, {0}, {1}) $,
$Q23=({{1/2}}, {{8/{15}}}, {-{1/2}}, {-{1/2}})$,
$Q24=({{1/ 2}}, {{8/{15}}}, {{1/ 2}}, {-{1/ 2}})$,
$Q25=({{1/2}}, {{8/{15}}}, {0}, {1})$,
$Q34=({0}, {0}, {1}, {0})$,
$Q35=({0}, {0}, {{1/ 2}}, {{3/ 2}})$,
$ Q45=({0}, {0}, {-{1/ 2}}, {{3/2}})$.
\label{Spectrum0}}
\begin{ruledtabular}
\begin{tabular}{c}
 Zero Modes for $\Sigma_i$ (in the $i$-th Row) \\
\hline\hline
$Q_X$:~ $\mathbf{(3, \bar 2)}_{Q12}$;
~$H_u'$:~ $\mathbf{(1, 2)}_{Q23}$;
~$S$:~$\mathbf{(1, 1)}_{Q45}$
~$\overline{D}_{\delta}$:~$\mathbf{(\bar 3, 1)}_{Q51}$ \\
\hline
$D_{\delta}$:~$\mathbf{(3, 1)}_{Q13}$;
~$S_2$:~$\mathbf{(1, 1)}_{Q34}$;
~$\overline{D}_X$:~$\mathbf{(\bar 3, 1)}_{Q41}$;
~~$H_d$:~$\mathbf{(1, \bar 2)}_{Q52}$ \\
\hline
 $H_u$:~ $\mathbf{(1, 2)}_{Q24}$;
~$S_1$:~$\mathbf{(1, 1)}_{Q53}$;
~~$S_1'$:~$\mathbf{(1, 1)}_{Q35}$;
~$H_d'$:~$\mathbf{(1, \bar 2)}_{Q42}$ \\
\end{tabular}
\end{ruledtabular}
\end{table}

The extra $U(1)$ gauge symmetries, $U(1)_{\alpha}
\times U(1)_{\beta} \times U(1)_{\gamma}$ 
can be broken at the compactification scale via Higgs mechanism, 
and thus the exotic quarks 
$Q_X$, $\overline{D}_X$, 
$D_{\delta}$, $\overline{D}_{\delta}$, and
the exotic doublets $H_u'$ and
$H_d'$ can acquire superheavy masses at this scale after these 
extra $U(1)$ gauge symmetries are broken. This can be achieved on the 3-brane at
the $Z_6$ fixed point, for example, $z=0$,
by introducing two exotic quarks $\overline{Q}'_X$
and $D'_{\delta}$ with quantum numbers 
$\mathbf{(\bar 3,  2)}_{{(5/6, 0, -1, 0)}}$ and
$\mathbf{(3, 1)}_{{(-1/3, 8/15, -1, 1)}}$ respectively
under the $SU(3)_C\times SU(2)_L\times U(1)_Y \times U(1)_{\alpha}
\times U(1)_{\beta} \times U(1)_{\gamma}$ gauge symmetry.
We also introduce a SM singlet Higgs field
${\widetilde S}_2$ which has the same quantum number as that of
$S_2$ and is localized on the 3-brane at $z=0$.
After ${\widetilde S}_2$ gets a VEV, the exotic quarks and Higgs
  doublets  can obtain the 
vector-like masses through the following brane-localized
superpotential,
\begin{equation}
W =  {\widetilde S}_2 H_u' H_d'  
+ {\widetilde S}_2 D_{\delta} \overline{D}_X
+ {\widetilde S}_2 Q_X \overline{Q}'_X
+ {\widetilde S}_2 D'_{\delta} \overline{D}_{\delta} \,.
\end{equation}
Similarly, $S_1'$ can be made superheavy.
Furthermore, the extra $U(1)$ symmetries can have gauge anomalies in 4D
since some of vector-like pairs of the zero modes are projected 
out by orbifolding.
Thus, we need to add brane-localized fields which has extra $U(1)$ charges 
to cancel the gauge anomalies,
and this can be an origin of the extra fields such as $\widetilde S_2$
and breaking of extra $U(1)$ symmetries.

After the $U(1)_{\alpha}
\times U(1)_{\beta} \times U(1)_{\gamma}$ gauge symmetry
is broken, we have the relevant superpotential
\begin{equation}
{\cal S} = \int d^6 x \left[ \int d^2 \theta \ g_6\left(
S H_u H_d - S S_1 S_2  \right)
 + {\rm H. C.}\right]\,.
\label{6D-NMSSM}
\end{equation}
Integrating out the two extra dimensions, we obtain,
at the GUT scale,
\begin{equation}
g_3~=~g_2~=~g_1~=~\lambda ~=~ \kappa ~=~ g_6/\sqrt{V} \,, 
\label{Couplings-U}
\end{equation}
where $V$ is the volume of extra dimensions, and
$g_3$, $g_2$ and $g_1\equiv {\sqrt {5/3}}\, g_Y$  are the 
4D gauge couplings for the SM gauge symmetry
$SU(3)_C$, $SU(2)_L$ and
$U(1)_Y$, respectively.
The true unification scale of the couplings is the cutoff scale ($M_*$)
in the orbifold models,
but for simplicity, we here assume that the compactification
scale is the GUT scale ($\sim 2 \times 10^{16}$ GeV)
so that the Higgs trilinear couplings $\lambda$ and $\kappa$ can be predicted.
We also neglect the brane-localized gauge kinetic
terms which are assumed to be suppressed by $\sqrt{V} M_*$ 
compared to the bulk kinetic term.

In our model, we introduce three families of the SM fermions on the
3-brane at the $Z_6$ fixed point $z=0$. Moreover,
we emphasize that the hypercharge interaction in the SM can be
one linear combination of the $U(1)_\alpha$ and $U(1)_Y$ in above $SU(8)$ model,
and then the hypercharge normalization may not be determined in this model
as in the usual orbifold GUT models when all the SM fermions are brane-localized fields.
However, if we identify $\overline D_\delta$ as a right-handed down-type quark field
in the presented choice of $Z_6$ charge assignment,
the hypercharge normalization in the SM can be the same as usual
$SU(5)$ normalization.

These additional zero modes such as exotic quarks, 
extra doublets and singlets can also be eliminated from the
zero modes of the compactified 4D spectrum by considering a
7-dimensional theory with $N=1$ supersymmetry and 
$SU(8)$ bulk gauge symmetry, and compactifying the three extra
dimensions on a $T^2/Z_6\times S^1/Z_2$ orbifold. Due to the orbifold
projections, the bulk $SU(8)$ gauge symmetry is broken directly down to 
the SM-like gauge symmetry, and there are
 only one pair of Higgs doublets and three SM  singlets in the Higgs sector
arising from the zero modes of bulk vector multiplet.
In this case, however, the hypercharge normalization is not determined completely.
When we consider larger gauge group, the quark and lepton fields 
can also be unified with the bulk gauge multiplet
and then the hypercharge normalization will be fixed naturally \cite{Gogoladze:2005zh}. 

\begin{figure}[t]
\includegraphics[viewport=0 0 700 500, clip, width=8cm]{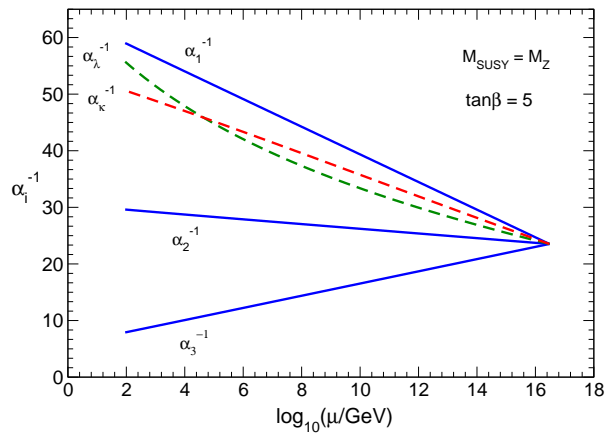}
\caption{ For $\tan\beta=5$, 
the unification of the SM gauge couplings ($\alpha_1, \alpha_2, \alpha_3$) 
and Higgs trilinear couplings ($\alpha_{\lambda}$ and $\alpha_{\kappa}$).} 
\label{gcu}
\end{figure}

The superpotential in Eq. (\ref{6D-NMSSM})
contains  five SM neutral complex scalar fields
and  (in the general case)
three phase symmetries in the scalar potential. One of these is the
$U(1)$ gauge symmetry related to the $Z$ boson, 
implying two unwanted global symmetries.
These will generally be spontaneously broken, implying two massless Goldstone
bosons. One of these has
large $H_d^0$ and $H_u^0$ components and is clearly 
excluded by the known experiment.
The second consists mainly of the $S$, $S_1$ and $S_2$  fields, and
is most likely also excluded, although a detailed
investigation is beyond the scope of this paper.
Let us give one possible solution to this problem. 
To break the $U(1)_{\beta} \times U(1)_{\gamma}$ gauge symmetry at GUT scale by
  Higgs mechanism, we introduce the SM singlet vector-like fields
$(N_1, \bar N_1)$, $(N_2, \bar N_2)$ and $(N_3, \bar N_3)$.
And the quantum numbers for $N_1$, $N_2$ and $N_3$ under
the $ U(1)_Y \times U(1)_{\alpha}
\times U(1)_{\beta} \times U(1)_{\gamma}$ gauge symmetry are
$(0,0,7/2, -21/2)$, $(0,0,-9,0)$  and $(0,0,4,12)$, respectively.
So, we can have the following non-renormalizable terms in the superpotential
\begin{eqnarray}
W^{\prime}=h_1\frac{S^7 N_1}{M_*^5}+ h_2\frac{S_2^9 N_2}{M_*^7}+ h_3\frac{S_1^8 N_3}{M_*^6}~,~\,
\end{eqnarray}
where  $h_1$, $h_2$,  $h_3$    are Yukawa coupling constants.
After $N_1$, $N_2$ and $N_3$ get VEVs, we obtain the effective
 superpotential at low scale
\begin{eqnarray}
W^{\prime}= h_1^{\prime}\frac{S^7}{M_*^4} + h_2^{\prime}\frac{S_1^9 }{M_*^6}+
 h_3^{\prime}\frac{S_2^8 }{M_*^5}~,~\,
\end{eqnarray}
where  $h_i^{\prime} = h_i \langle N_i \rangle/M_*$.
As shown in Ref. \cite{Abel:1995wk}, these non-renormalizable terms in
 the above superpotential do not  generate the dangerous quadratically divergent 
tadpoles for the SM singlet fields $S$ and $S_i$ \cite{LAX}.
Also,  we do not have   global symmetries in our model, so,
there are no massless Goldstone bosons and
 there is no domain wall problem after the Higgs doublets and SM
singlets obtain VEVs.

{\em Phenomenology --}
We now briefly discuss the phenomenological implication of our model,
in particular, the implication of the effective superpotential
given by Eq.~(\ref{6D-NMSSM}), and the unification of Higgs trilinear couplings
with the SM gauge couplings, Eq.~(\ref{Couplings-U}). 
The unification prediction can be tested
by using the appropriate renormalization group equations (RGEs) for these couplings.
For numerical calculations, we consider two-loop RGE runnings 
for the SM gauge couplings and top quark Yukawa coupling $y_t$,
 and one-loop RGE runnings 
for the Higgs trilinear couplings ($\lambda$ and $\kappa$) 
\cite{King:1995ys}, with
conversion from $\overline{MS}$ scheme to dimensional reduction
$(\overline{DR})$ scheme. We also include the standard supersymmetric
threshold corrections at low energy by choosing a single scale
$M_{SUSY}=M_Z$ where $M_Z$ is the $Z$-boson
mass \cite{Langacker:1992rq}. The relevant RGEs are
\begin{eqnarray}
\frac{d \alpha_{\lambda}}{dt} &=&\frac{\alpha_{\lambda}}{2\pi}
(\alpha_{\kappa}+4\alpha_{\lambda}+3 \alpha_t- \frac{3}{5}\alpha_1 - 3\alpha_2)\,, 
\label{RGE-lambda}\\
\frac{d \alpha_{\kappa}}{dt} &=&
\frac{\alpha_{\kappa}}{2\pi}(3\alpha_{\kappa}+2\alpha_{\lambda})\,, \\
\frac{d \alpha_{t}}{dt} &=& \left[ \frac{d \alpha_{t}}{dt}\right]  _{\rm MSSM} 
\nonumber \\
&& + \frac{\alpha_{t}}{2\pi} \left( \alpha_{\lambda} 
 -\frac{1}{4\pi}\alpha_{\lambda}
\left( 3\alpha_t+ 3\alpha_{\lambda}+\alpha_{\kappa}\right) \right), \\
\frac{d \alpha_{2}}{dt} &=& \left[ \frac{d \alpha_{2}}{dt}\right] _{\rm MSSM}
+ \frac{\alpha_{2}^2}{8\pi^2}\left( -2\alpha_{\lambda}\right) \,, \\
\frac{d \alpha_{1}}{dt} &=& \left[ \frac{d \alpha_{1}}{dt}\right] _{\rm MSSM}
+ \frac{\alpha_{1}^2}{8\pi^2}
\left( -{6\over 5}\alpha_{\lambda}\right) \,,\, \label{RGE-alpha1}
\end{eqnarray}
where $t$ is the ${\rm log}$ of renormalization scale, 
$\alpha_i=g^2_i/(4\pi)$, $\alpha_{\lambda} = \lambda^2/(4\pi)$,
$\alpha_{\kappa} =\kappa^2/(4\pi)$, $\alpha_{t} =y_t^2/(4\pi)$,
and  the bracket $\left[\, \right]$ denotes the corresponding
two-loop RGEs in the MSSM.
We use the values of SM gauge couplings at $M_Z$ in Ref.~\cite{pdg}
and the top quark mass to be 178 GeV.

\begin{figure}[t]
\includegraphics[viewport=0 0 700 500, clip, width=8cm]{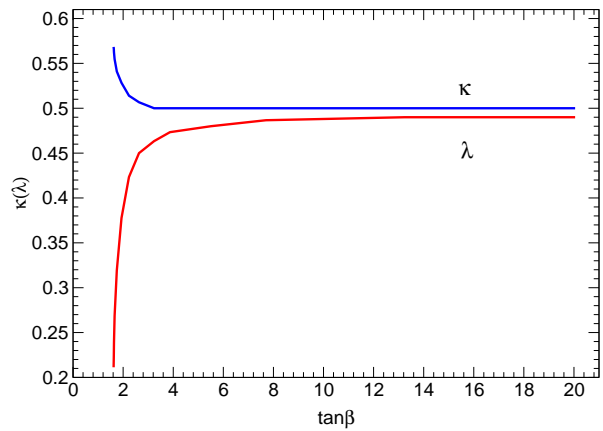}
\caption{ The  Higgs trilinear couplings $\lambda$ and $\kappa$
at the weak scale versus $\tan\beta$. }
\label{Coupling-LK}
\end{figure}

\begin{figure}
\includegraphics[viewport=80 15 580 722, clip, width=5.7cm, angle=270]{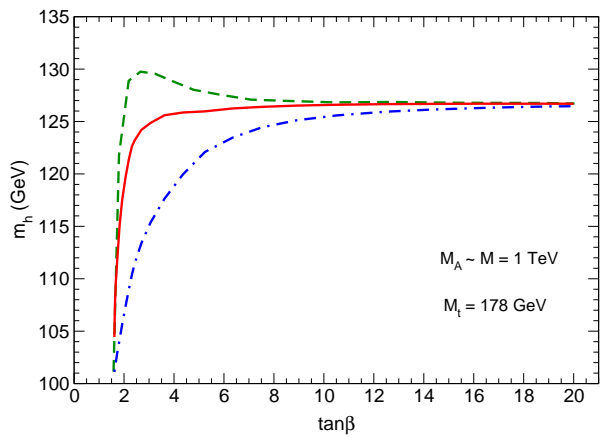}
\caption{ The upper bounds on
the lightest CP-even neutral Higgs $h^0$ mass
in the MSSM, the NMSSM and our model versus $\tan\beta$:
the blue dash-dot line, green dash line, and 
red solid line corresponds to the MSSM, the NMSSM, and our model,
respectively. }
\label{Higgs-Mass}
\end{figure}

Our results for the unification of the gauge and  Higgs trilinear
couplings using the RGEs, 
Eqs.~(\ref{RGE-lambda}-\ref{RGE-alpha1}), are shown in Fig.~\ref{gcu}
for $\tan\beta=5$ where 
$\tan\beta \equiv \langle H_u^0 \rangle/\langle H_d^0 \rangle$.
Also the predictions for the couplings $\lambda$ and $\kappa$ at
the weak scale, $M_Z$, for various values of $\tan\beta$ are
shown in Fig.~\ref{Coupling-LK}. 

Since the values of $\lambda$ and $\kappa$ are predicted in our 
model, we can calculate the upper bound on the mass of the
lightest CP-even neutral Higgs boson $h^0$  by using  the full one-loop  
and the leading  logarithmic 
two-loop corrections \cite{Okada:1990vk}. In Fig.~\ref{Higgs-Mass},
we plot the upper bounds for
the $h^0$ mass
in the MSSM, the NMSSM and our model versus $\tan\beta$. Note that for a large range
of $\tan\beta$, 
the mass bound in our model is larger than in the MSSM, but less than in the 
 NMSSM in which the values of $\lambda$ and $\kappa$ are arbitrary. 
 For this calculation we use the approximation that the mass of CP-odd Higgs ($M_A$)
 is order of the square root of the 
arithmetic average of the stop squared-mass eigenvalues ($M$).
The validity of our prediction can be tested in the upcoming LHC.

{\em Conclusions --}
We have presented a supersymmetric Standard Model in which the
Higgs trilinear couplings $\lambda$ and $\kappa$ are unified with
the three SM gauge couplings at the unification scale.
This is an orbifold GUT model in 6D with $N=2$
supersymmetry. The symmetry is broken down to the SM
gauge symmetry in four dimensions via orbifold compactification
as well as via Higgs mechanism. The unification prediction is
in good agreement with experiments. The predicted upper bound for the lightest
CP-even Higgs mass is somewhat larger than in the MSSM, and can be tested
at the LHC. The detail model buildings which also include the possible unification
of the third-family Yukawa couplings, and their phenomenological
consequences will be presented elsewhere.

{\em Acknowledgments --}
This research  was supported in part by the
  National Science Foundation under Grant No.
PHY-0098791 (IG) and No.~PHY-0070928 (TL), by 
the Natural Sciences and Engineering Research Council of Canada (YM),
and by the Department of Energy 
Grants ~DE-FG02-04ER46140 and DE-FG02-04ER41306 (SN).



\begin{thebibliography}{99}



\bibitem{Langacker:1990jh}
  S.~Dimopoulos and H.~Georgi,
  Nucl.\ Phys.\ B {\bf 193}, 150 (1981);
%
  S.~Dimopoulos, S.~Raby and F.~Wilczek,
  Phys.\ Rev.\ D {\bf 24}, 1681 (1981);
%
  N.~Sakai,
  Z.\ Phys.\ C {\bf 11}, 153 (1981);
%
  L.~E.~Ibanez and G.~G.~Ross,
  Phys.\ Lett.\ B {\bf 105}, 439 (1981);
%
  M.~B.~Einhorn and D.~R.~T.~Jones,
  Nucl.\ Phys.\ B {\bf 196}, 475 (1982);
%
  W.~J.~Marciano and G.~Senjanovic,
  Phys.\ Rev.\ D {\bf 25}, 3092 (1982).
%

\bibitem{Amaldi:1991cn}
U.~Amaldi, W.~de Boer and H.~Furstenau,
Phys.\ Lett.\ B {\bf 260}, 447 (1991);
%
  J.~R.~Ellis, S.~Kelley and D.~V.~Nanopoulos,
  Phys.\ Lett.\ B {\bf 249}, 441 (1990);
%
P.~Langacker and M.~X.~Luo,
Phys.\ Rev.\ D {\bf 44}, 817 (1991);
%
C.~Giunti, C.~W.~Kim and U.~W.~Lee,
Mod.\ Phys.\ Lett.\ A {\bf 6} (1991) 1745.




\bibitem{Okada:1990vk}
Y.~Okada, M.~Yamaguchi and T.~Yanagida,
Prog.\ Theor.\ Phys.\  {\bf 85}, 1 (1991);
%
J.~R.~Ellis, G.~Ridolfi and F.~Zwirner,
Phys.\ Lett.\ B {\bf 257}, 83 (1991);
%
H.~E.~Haber and R.~Hempfling,
Phys.\ Rev.\ Lett.\  {\bf 66}, 1815 (1991);
M.~Carena, J.~R.~Espinosa, M.~Quiros and C.~E.~M.~Wagner,
Phys.\ Lett.\ B {\bf 355}, 209 (1995).




\bibitem{Nilles:1982dy}
P.~Fayet,
Nucl.\ Phys.\ B {\bf 90}, 104 (1975);
H.~P.~Nilles, M.~Srednicki and D.~Wyler,
Phys.\ Lett.\ B {\bf 120}, 346 (1983);
%
J.~M.~Frere, D.~R.~T.~Jones and S.~Raby,
Nucl.\ Phys.\ B {\bf 222}, 11 (1983);
%
J.~P.~Derendinger and C.~A.~Savoy,
Nucl.\ Phys.\ B {\bf 237}, 307 (1984).

\bibitem{Orbifold}
Y.~Kawamura,
Prog.\ Theor.\ Phys.\  {\bf 103}, 613 (2000);
%
{\it ibid.} {\bf 105}, 999 (2001); {\it ibid.} {\bf 105}, 691 (2001);
G. Altarelli and F. Feruglio, 
Phys.\ Lett.\ B {\bf 511}, 257 (2001);
A.~B.~Kobakhidze,
Phys.\ Lett.\ B {\bf 514}, 131 (2001);
L. Hall and Y. Nomura, Phys.\ Rev.\ D {\bf 64}, 055003 (2001);
A. Hebecker and J. March-Russell, 
Nucl.\ Phys.\ B {\bf 613}, 3 (2001);
 T. Li, Phys.\ Lett.\ B {\bf 520}, 377 (2001);
Nucl.\ Phys.\ B {\bf 619}, 75 (2001).

\bibitem{GMS}
I.~Gogoladze, Y.~Mimura and S.~Nandi,
Phys.\ Lett.\ B {\bf 562}, 307 (2003);
Phys.\ Rev.\ Lett.\  {\bf 91}, 141801 (2003);
Phys.\ Rev.\ D {\bf 69}, 075006 (2004);
I.~Gogoladze, Y.~Mimura, S.~Nandi and K.~Tobe,
Phys.\ Lett.\ B {\bf 575}, 66 (2003);
T.~Kobayashi, S.~Raby and R.~J.~Zhang,
Nucl.\ Phys.\ B {\bf 704}, 3 (2005).


\bibitem{NMASWS}
N.~Marcus, A.~Sagnotti and W.~Siegel,
Nucl.\ Phys.\ B {\bf 224}, 159 (1983);
N.~Arkani-Hamed, T.~Gregoire and J.~Wacker,
JHEP {\bf 0203}, 055 (2002).

\bibitem{Li:2003ee}
T.~Li,
JHEP {\bf 0403}, 040 (2004).


\bibitem{Gogoladze:2005zh}
  I.~Gogoladze, T.~Li, Y.~Mimura and S.~Nandi,
  hep-ph/0504082.




\bibitem{Abel:1995wk}
  S.~A.~Abel, S.~Sarkar and P.~L.~White,
  Nucl.\ Phys.\ B {\bf 454}, 663 (1995);
  S.~A.~Abel,
  Nucl.\ Phys.\ B {\bf 480}, 55 (1996).




\bibitem{LAX}  
  H.~P.~Nilles, M.~Srednicki and D.~Wyler,
  Phys.\ Lett.\ B {\bf 124}, 337 (1983);
%
  A.~B.~Lahanas,
  Phys.\ Lett.\ B {\bf 124}, 341 (1983);
%
  U.~Ellwanger,
  Phys.\ Lett.\ B {\bf 133}, 187 (1983);
%
  J.~Bagger and E.~Poppitz,
  Phys.\ Rev.\ Lett.\  {\bf 71}, 2380 (1993);
%
  J.~Bagger, E.~Poppitz and L.~Randall,
  Nucl.\ Phys.\ B {\bf 455}, 59 (1995);
%
  V.~Jain,
  Phys.\ Lett.\ B {\bf 351}, 481 (1995);
%
  C.~Panagiotakopoulos and K.~Tamvakis,
  Phys.\ Lett.\ B {\bf 446}, 224 (1999).





\bibitem{King:1995ys}
S. P. Martin and M. T. Vaughn, Phys. Rev. D {\bf 50}, 2282 (1994);
  S.~F.~King and P.~L.~White,
  Phys.\ Rev.\ D {\bf 52}, 4183 (1995).


\bibitem{Langacker:1992rq}
P.~Langacker and N.~Polonsky,
Phys.\ Rev.\ D {\bf 47}, 4028 (1993);
M.~Carena, S.~Pokorski and C.~E.~M.~Wagner,
Nucl.\ Phys.\ B {\bf 406}, 59 (1993).

\bibitem{pdg}
S.~Eidelman {\it et al.}  [Particle Data Group Collaboration],
Phys.\ Lett.\ B {\bf 592}, 1 (2004).









\end{thebibliography}
\end{document}